\documentclass[10pt, conference, compsocconf]{IEEEtran}
\usepackage{graphicx}
\usepackage{url}
\usepackage{amsmath}

\usepackage{algorithm}
\usepackage[noend]{algorithmic}

\ifCLASSINFOpdf
  
\else
 
\fi

\graphicspath{{images//}}

\begin{document}
%
\title{SLPA: Uncovering Overlapping Communities in Social Networks via A Speaker-listener Interaction Dynamic Process}

\author{\IEEEauthorblockN{Jierui Xie and Boleslaw K. Szymanski}
\IEEEauthorblockA{Department of Computer Science\\
Rensselaer Polytechnic Institute\\
Troy, New York 12180\\
Email: \{xiej2, szymansk\}@cs.rpi.edu}
\and
\IEEEauthorblockN{Xiaoming Liu}
\IEEEauthorblockA{Department of Computer Science\\
University of the Western Cape\\
Belville, South Africa\\
Email: andyliu5738@gmail.com}
}

\maketitle

\begin{abstract}
\textit{Overlap} is one of the characteristics of social networks, in which a person may belong to more than one social group. For this reason, discovering overlapping structures is necessary for realistic social analysis. In this paper, we present a novel, general framework to detect and analyze both individual overlapping nodes and entire communities. In this framework, nodes exchange labels according to dynamic interaction rules. A specific implementation called Speaker-listener Label Propagation Algorithm (SLPA\footnote{SLPA 1.0: \url{   https://sites.google.com/site/communitydetectionslpa/}}) demonstrates an excellent performance in identifying both overlapping nodes and overlapping communities with different degrees of diversity.

\end{abstract}


\IEEEpeerreviewmaketitle

\section{Introduction}
\label{sec:intr}
Modular structure is considered to be the building block of real-world networks as it often accounts for the functionality of the system. It has been well understood that people in a social network are naturally characterized by \textit{multiple} community memberships. For example, a person usually has connections to several social groups like family, friends and colleges; a researcher may be active in several areas; in the Internet, a person can simultaneously subscribe to an arbitrary number of groups. 

For this reason, overlapping community detection algorithms have been investigated. These algorithms aim to discover a \textit{cover} \cite{LancichinettiNMI:2009}, which is defined as a set of clusters in which each node belongs to at least one cluster. In this paper, we propose an efficient algorithm to identify both individual overlapping nodes and the entire overlapping communities using the underlying network structure alone. 

\section{Related Work}
\label{sec:rw}
The work on detecting overlapping communities was previously proposed by Palla \cite{CPM:2005} with the clique percolation algorithm (CPM). CPM is based on the assumption that a community consists of fully connected subgraphs and detects overlapping communities by searching for each such subgraph for \textit{adjacent} cliques that share with it at least certain number of nodes. CPM is suitable for networks with dense connected parts. 

Another line of research is based on maximizing a local benefit function. Baumes \cite{Baumes1:2005} proposed the iterative scan algorithm (IS). IS expands seeded small cluster cores by adding or removing nodes until the local density function cannot be improved. The quality of discovered communities depends on the quality of seeds. LFM \cite{LFM:2009} expands a community from a random seed node until the fitness function is locally maximal. LFM depends significantly on a parameter of the fitness function that controls the size of the communities.  

CONGA \cite{CONGA:2007} extends Girvan and Newman's divisive clustering algorithm by allowing a node to split into multiple copies. Both splitting betweenness defined based on the number of shortest paths on the imaginary edge and the usual edge betweenness are considered. In the refined version of CONGO \cite{CONGO:2008}, local betweenness is used  to optimize the speed.

Copra \cite{COPRA:2010} is an extension of the label propagation algorithm \cite{Raghavan:2007} for overlapping community detection. Each node updates its \textit{belonging coefficients} by \textit{averaging} the coefficients over all its neighbors. Copra produces a number of small size communities in some networks.

EAGLE \cite{EAGLE:2009} uses the agglomerative framework to produce a dendrogram. All maximal cliques that serve as initial communities are first computed. Then, the pair of communities with maximum similarity is merged iteratively. Expensive computation is one drawback of this algorithm.

Fuzzy clustering has also been extended to overlapping community detection. Zhang \cite{ZhangFuzzy:2007} used the spectral method to embed the graph into \textit{k}-1 dimensional Euclidean space. Nodes are then clustered by the fuzzy c-mean algorithm. Nepusz \cite{Nepusz:2008} modeled the overlapping community detection as a nonlinear constraint optimization problem. Psorakis et al. \cite{NMF:2010} proposed a model based on Bayesian nonnegative matrix factorization (NMF). 

The idea of partitioning links instead of nodes to discover community structure has also been explored \cite{YYLinkClustering:2010,LineGraph1:2010}. As a result, the node partition of a line (or link) graph leads to an edge partition of the original graph.

\section{SLPA: Speaker-listener Label Propagation Algorithm}
\label{sec:SLPA}
The algorithm proposed in this paper is an extension of the Label Propagation Algorithm (LPA) \cite{Raghavan:2007}. In LPA, each node holds only a single label that is iteratively updated by adopting the majority label in the neighborhood. Disjoint communities are discovered when the algorithm converges. One way to account for overlap is to allow each node to possess multiple labels as proposed in \cite{Gregory:2010}. Our algorithm follows this idea but applies different dynamics with more general features.

In the dynamic process, we need to determine 1) how to spread nodes' information to others; 2) how to process the information received from others. The critical issue related to both questions is how information should be \textit{maintained}. We propose  a speaker-listener based information propagation process (SLPA) to mimic human communication behavior. 

In SLPA, each node can be a listener or a speaker. The roles are switched depending on whether a node serves as an information provider or information consumer. Typically, a node can hold as many labels as it likes, depending on what it has experienced in the stochastic processes driven by the underlying network structure. A node \textit{accumulates} knowledge of repeatedly observed labels instead of erasing all but one of them. Moreover, the more a node observes a label, the more likely it will spread this label to other nodes (mimicking people's preference of spreading most frequently discussed opinions). 

In a nutshell, SLPA consists of the following three stages (see algorithm \ref{alg1} for pseudo-code):

\begin{description}
\item[1)] First, the memory of each node is initialized with this node's id (i.e., with a unique label).
\item[2)] Then, the following steps are repeated until the stop criterion is satisfied:
	\begin{description}
	\item[a.] One node is selected as a listener.
	\item[b.] Each neighbor of the selected node sends out a \textit{single} label following certain \textit{speaking rule}, such as selecting a random label from its memory with probability proportional to the occurrence frequency of this label in the memory.
	\item[c.] The listener accepts \textit{one} label from the collection of labels received from neighbors following certain \textit{listening rule}, such as selecting the most popular label from what it observed in the \textit{current} step.
	\end{description}
\item[3)] Finally, the post-processing based on the labels in the memories of nodes is applied to output the communities.
\end{description}
\begin{algorithm}
\caption[caption]{: SLPA(\textit{T}, \textit{r})}
\label{alg1}
\begin{algorithmic}
\STATE [n,Nodes]=loadnetwork();
\STATE \textit{Stage 1: initialization}
	\FOR{$i=1:n$}  
		\STATE Nodes(i).Mem={i};
	\ENDFOR
\STATE \textit{Stage 2: evolution}
	\FOR{$t=1:T$}
	    \STATE Nodes.ShuffleOrder();
	    \FOR{$i=1:n$}
	        \STATE Listener=Nodes(i);
	        \STATE Speakers=Nodes(i).getNbs();
	        \FOR{$j=1:Speakers.len$ }
	             \STATE LabelList(j)= Speakers(j).speakerRule();
	        \ENDFOR
	        \STATE w=Listener.listenerRule(LabelList);
	        \STATE Listener.Mem.add(w); 
	    \ENDFOR
	\ENDFOR
\STATE \textit{Stage 3: post-processing}
	\FOR{$i=1:n$}
	    \STATE remove Nodes(i) labels seen with probability $< r$;
	\ENDFOR
\end{algorithmic}
\end{algorithm}
SLPA utilizes an asynchronous update scheme, i.e., when updating a listener's memory at time $t$, some already updated neighbors have memories of size $t$ and some other neighbors still have memories of size $t-1$. SLPA reduces to LPA when the size of memory is limited to one and the stop criterion is convergence of all labels.

It is worth noticing that each node in our system has a \textit{memory} and takes into account information that has been observed in the \textit{past} to make current decision. This feature is typically absent in other label propagation algorithms such as \cite{Gregory:2010,JieruiXieLPA:2010}, where a node updates its label completely forgetting the old knowledge. This feature allows us to combine the accuracy of the asynchronous update with the stability of the synchronous update \cite{Raghavan:2007}. As a result, the fragmentation issue of producing a number of small size communities observed in Copra in some networks, is avoided.
\subsection{Stop Criterion} The original LPA stop criterion of having every node assigned the most popular label in its neighborhood does not apply to the multiple labels case. Since neither the case where the algorithm reaches a single community (i.e., a special convergence state) nor the oscillation (e.g., on bipartite network) would affect the stability of SLPA, we can stop at any time as long as we collect sufficient information for post-processing. In the current implementation, SLPA simply stops when the predefined maximum number of iterations \textit{T} is reached. In general, SLPA produces relatively stable outputs, independent of network size or structure, when \textit{T} is greater than 20. 
\begin{figure*}[th]
\centering	 
	 \begin{minipage}[t]{0.45\linewidth}
	 \centering	
			\includegraphics[scale=0.6]{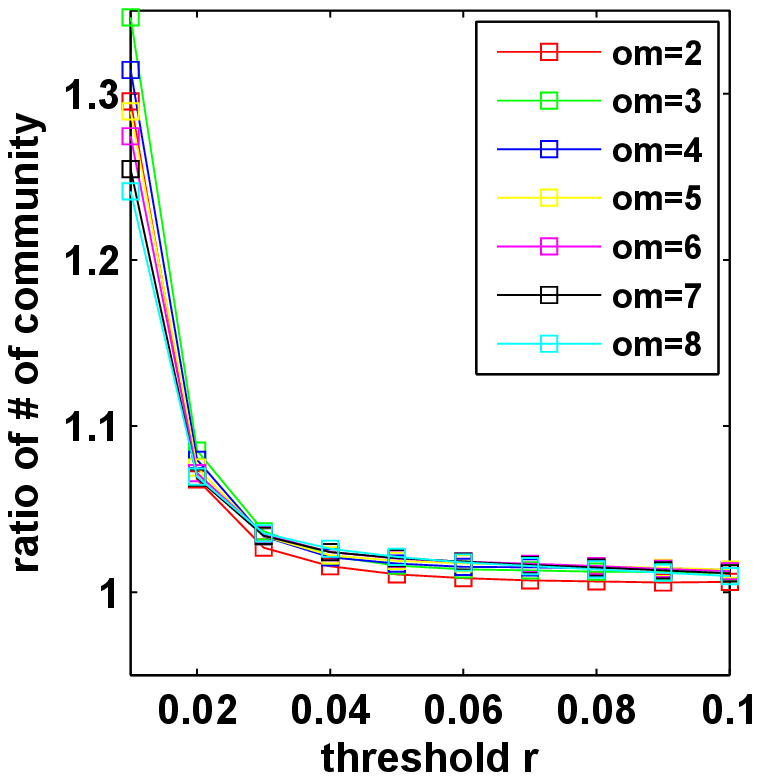}
			\caption{The convergence behavior of the parameter in LFR benchmarks with $n=5000$. y-axis is the ratio of the numbers of detected to true communities.}
			\label{fig:SLPARatioNumCom}
	 \end{minipage}	
	 \hspace{0.1cm}  
	 \begin{minipage}[t]{0.45\linewidth}
	 \centering	
	 	\includegraphics[scale=0.7]{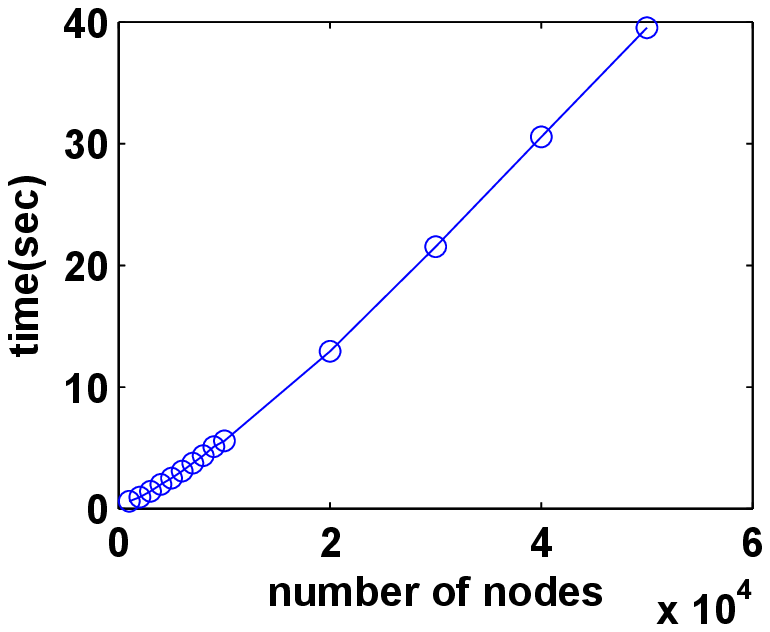}
			\caption{The execution time in second of SLPA in LFR benchmarks with $\overline{k}=20$. $n$ ranges from 1000 to 50000.}
			\label{fig:executionTime}	 
			\end{minipage}
\end{figure*}		 			 	
\subsection{Post-processing and Community Detection} SLPA collects only label information that reflects the underlying network structure during the evolution. The detection of communities is performed when the stored information is post-processed. Given the memory of a node, SLPA converts it into a probability distribution of labels. This distribution defines the \textit{strength} of association to  communities to which the node belongs. This distribution can be used for \textit{fuzzy communities} detection \cite{SteveSurvey:2011}. More often than not, one would like to produce \textit{crisp} communities in which the membership of a node to a given community is \textit{binary}, i.e., either a node is in a community or not. To this end, a simple thresholding procedure is performed. If the probability of seeing a particular label during the whole process is less than a given threshold $r\in [0,1]$, this label is deleted from a node's memory. After thresholding, connected nodes having a particular label are grouped together and form a community. If a node contains multiple labels, it belongs to more than one community and is therefore called an \textit{overlapping node}. In SLPA, we remove nested communities, so the final communities are \textit{maximal}.

As shown in Fig. \ref{fig:SLPARatioNumCom}, SLPA converges (i.e., producing similar output) quickly as the parameter $r$ varies. The effective range is typically narrow. Note that the threshold is used only in the post-processing. It means that the dynamics of SLPA is completely determined by the network structure and the interaction rules. The number of memberships is constrained only by the node degree. In contrast, Copra uses a parameter to control the \textit{maximum} number of memberships granted during the iterations.
\subsection{Complexity}
The initialization of labels requires $O(n)$, where $n$ is the total number of nodes. The outer loop is controlled by the user defined maximum iteration $\textit{T}$,  which is a small constant\footnote{In our experiments, we used $\textit{T}$ set to 100.}. The inner loop is controlled by $n$. Each operation of the inner loop executes one speaking rule and one listening rule. For the speaking rule, selecting a label from the memory proportionally to the frequencies is, in principle, equivalent to randomly selecting an element from the array, which is $O(1)$ operation. For listening rule, since the listener needs to check all the labels from its neighbors, it takes $O(K)$ on average, where $K$ is the average degree. The complexity of the dynamic evolution (i.e., stage 1 and 2) for the asynchronous update is $O(Tm)$ on an arbitrary network and $O(Tn)$ on a sparse network, when $m$ is the total number of edges. In the post-processing, the thresholding operation requires $O(Tn)$ operations since each node has a memory of size $\textit{T}$. 

Therefore, the time complexity of the entire algorithm is $O(Tn)$ in sparse networks. For a naive implementation, the execution time on synthetic networks (see section \ref{sec:Exp}) scales slightly faster than a linear growth with $n$ as shown in Fig. \ref{fig:executionTime}. 

\begin{figure*}[!th]
\centering	  
	 \begin{minipage}[t]{0.45\linewidth}
	 \centering	  
			\includegraphics[scale=0.6]{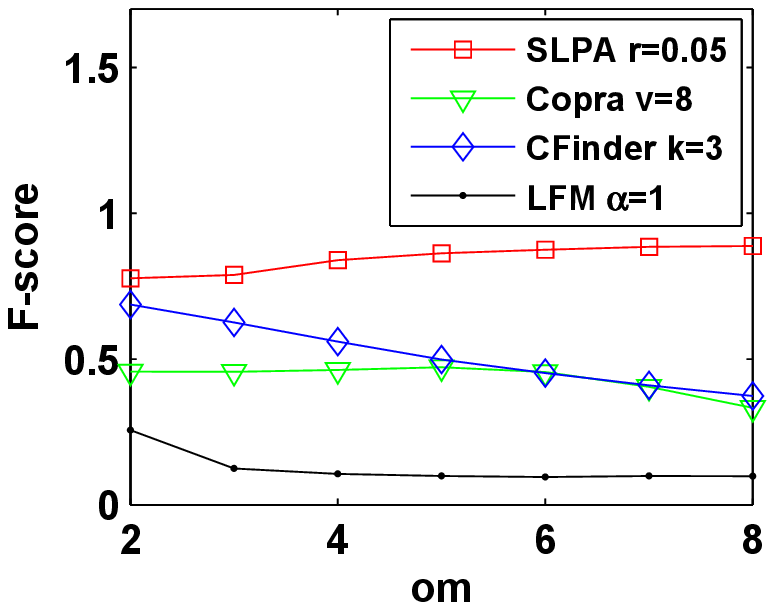}
			\caption{F-score for networks with $n=5000$, $\overline{k}=10$, $\mu=0.1$. }
			\label{fig:FN5000Bmu01}
	 \end{minipage}	
	 \hspace{0.1cm}  
	 	 \begin{minipage}[t]{0.45\linewidth}
	 	 \centering	  
			\includegraphics[scale=0.6]{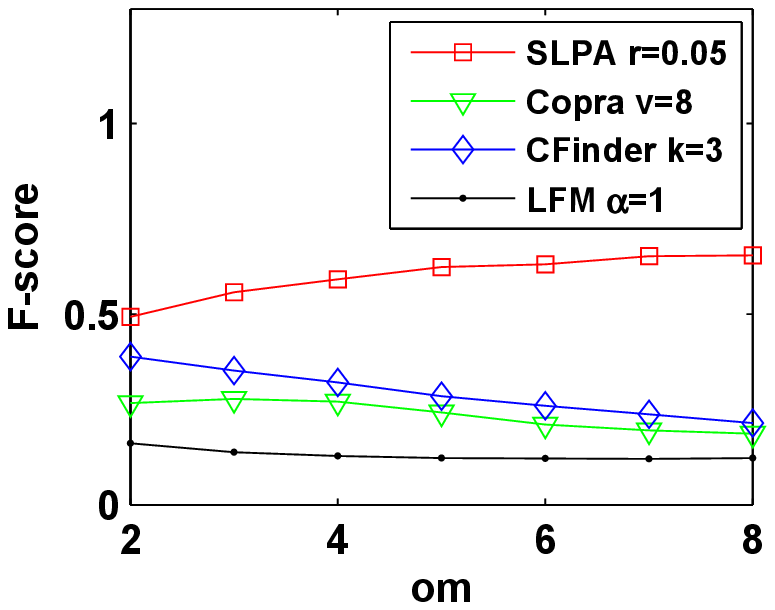}
			\caption{F-score for networks with $n=5000$, $\overline{k}=10$, $\mu=0.3$.}
	 		\label{fig:FN5000Bmu03}
	 \end{minipage}		 
\end{figure*}

\begin{figure*}[!th]
\centering
	 \begin{minipage}[t]{0.3\linewidth}
			\includegraphics[scale=0.65]{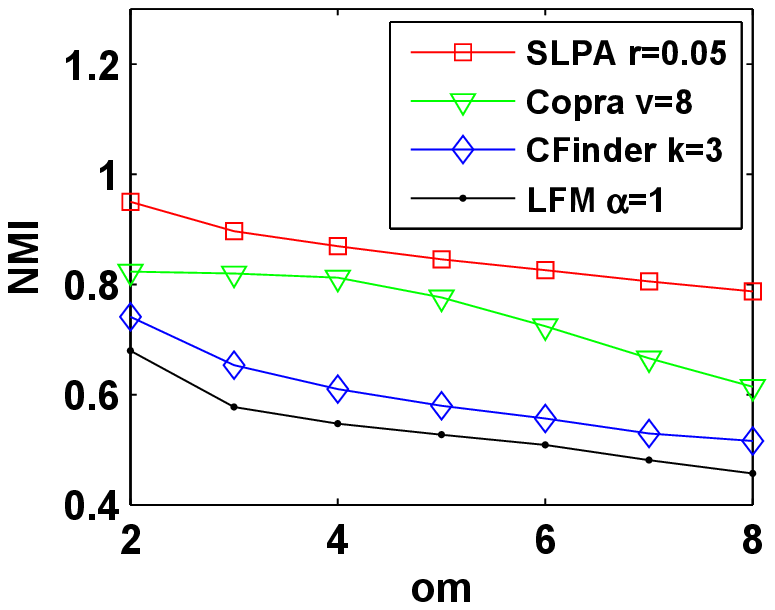}
			\caption{NMI for networks with $n=5000$, $\overline{k}=10$, $\mu=0.1$.}
			\label{fig:NMIN5000Bmu01}
	 \end{minipage}	
	 \hspace{0.2cm}  
	 \begin{minipage}[t]{0.3\linewidth}
			\includegraphics[scale=0.65]{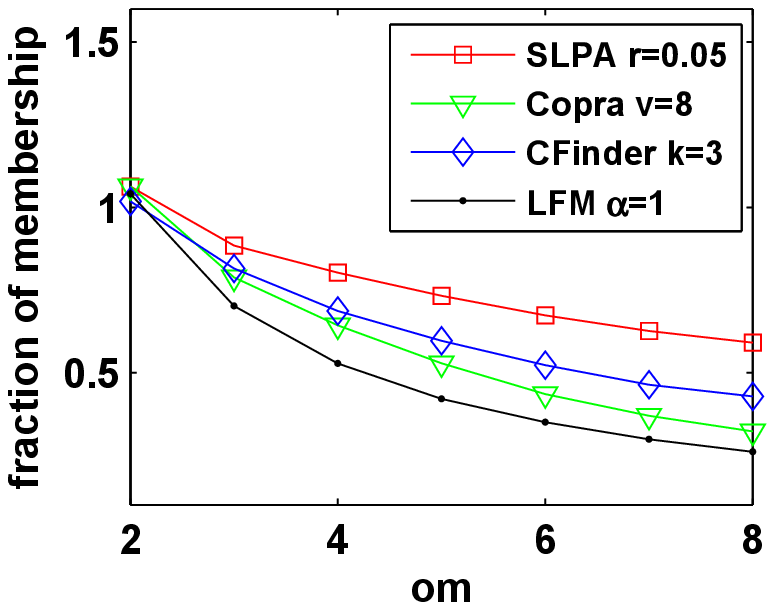}
			\caption{Ratio of the detected to the known numbers of memberships for networks with $n=5000$, $\overline{k}=10$, $\mu=0.1$. Values over 1 are possible when more memberships are detected than there are known to exist.}
			\label{fig:framembershipN5000Bmu01}
	 \end{minipage}	
	 \hspace{0.2cm}  
	 \begin{minipage}[t]{0.3\linewidth}
			\includegraphics[scale=0.65]{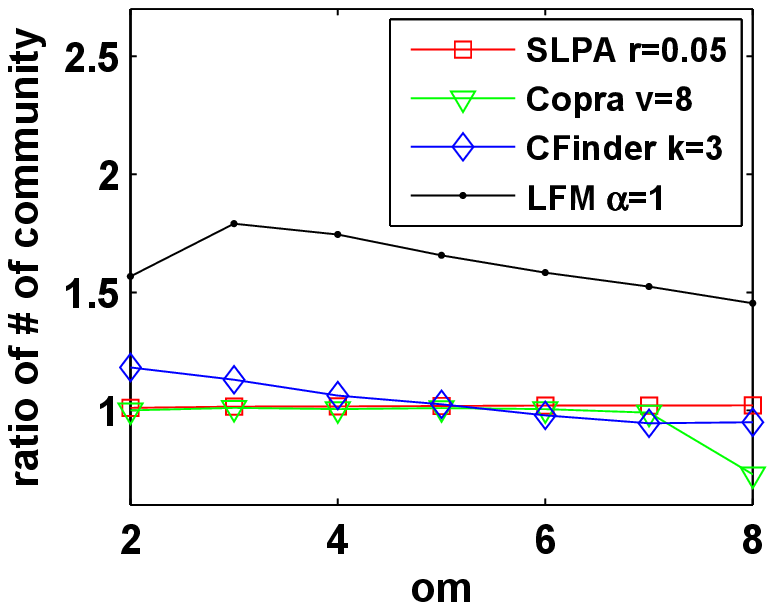}
			\caption{Ratio of the detected to the known numbers of communities for networks with $n=5000$, $\overline{k}=10$, $\mu=0.1$. Values over 1 are possible when more communities are detected than there are known to exist.}
			\label{fig:rationumcomN5000Bmu01}
	 \end{minipage}	
	\end{figure*} 
	  
\begin{figure*}[!th]
\centering
	 \begin{minipage}[t]{0.3\linewidth}
			\includegraphics[scale=0.65]{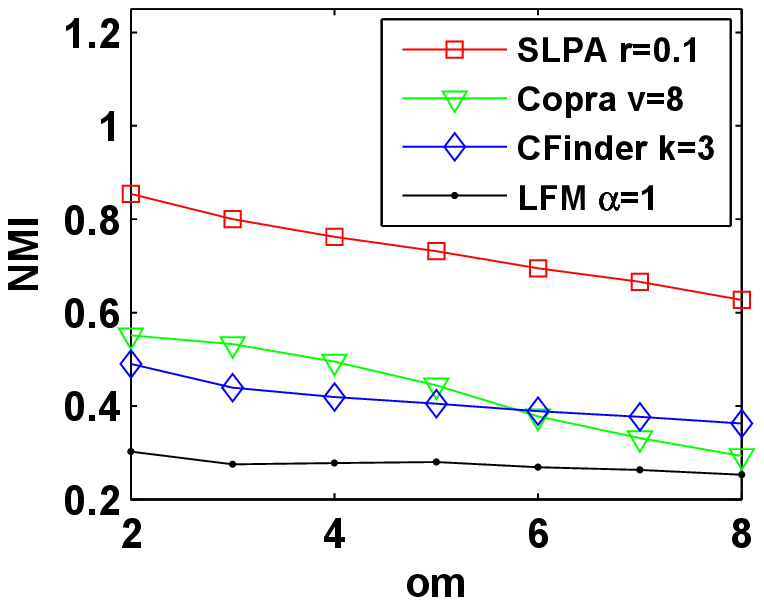}
			\caption{NMI for networks with $n=5000$, $\overline{k}=10$, $\mu=0.3$.}
	 		\label{fig:NMIN5000Bmu03}
	 \end{minipage}
	 \hspace{0.2cm}  
	 \begin{minipage}[t]{0.3\linewidth}
			\includegraphics[scale=0.65]{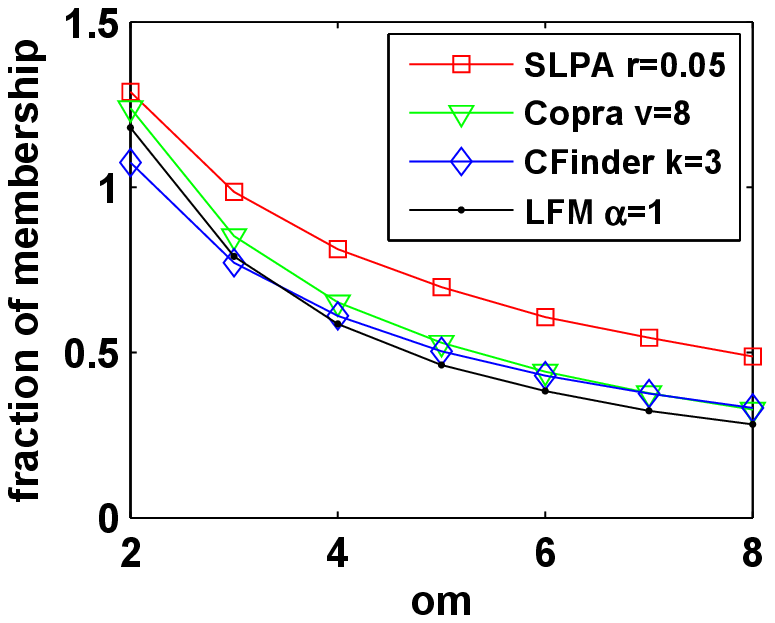}
			\caption{Ratio of the detected to the known numbers of memberships for networks with $n=5000$, $\overline{k}=10$, $\mu=0.3$. Values over 1 are possible when more memberships are detected than there are known to exist.}
	 		\label{fig:framembershipN5000Bmu03}
	 \end{minipage}
	 \hspace{0.2cm}  
	 \begin{minipage}[t]{0.3\linewidth}
			\includegraphics[scale=0.65]{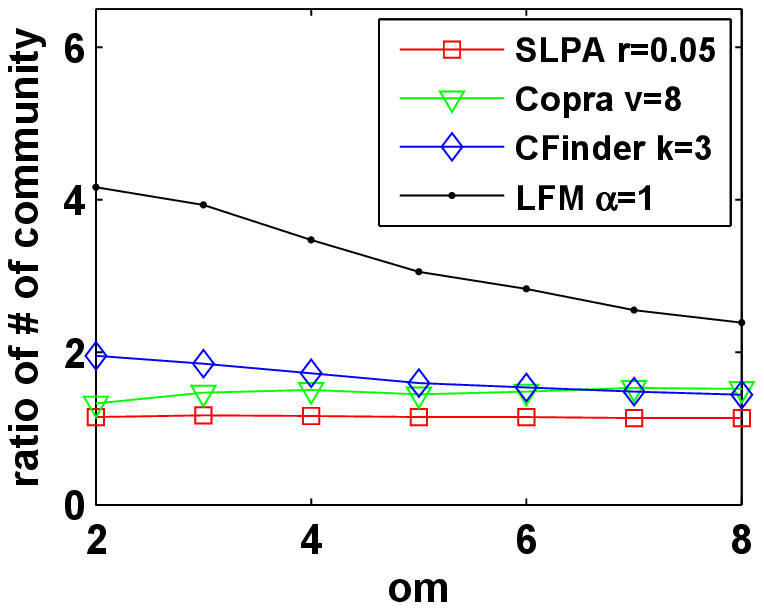}
			\caption{Ratio of the detected to the known numbers of communities for networks with $n=5000$, $\overline{k}=10$, $\mu=0.3$. Values over 1 are possible when more communities are detected than there are known to exist.}
	 		\label{fig:rationumcomN5000Bmu03}
	 \end{minipage}	
\end{figure*}
\section{Experiments and Results}
\label{sec:Exp}

\subsection{Benchmark Networks} 
To study the behavior of SLPA for overlapping community detection, we conducted extensive experiments on both synthetic and real-world networks. Table \ref{table:real1} lists the classical social networks for our tests and their statistics \footnote{Data are available at \url{http://www-personal.umich.edu/~mejn/netdata/} and  \url{http://deim.urv.cat/~aarenas/data/welcome.htm}}. For synthetic networks, we adopted the LFR benchmark \cite{LFR:2008}, which is a special case of the planted $l$-partition model, but characterized by heterogeneous distributions of node degrees and community sizes. 

In our experiments, we used networks with size $n=5000$. The average degree is kept at $\overline{k}=10$ which is of the same order as most of the real-world networks we tested. The rest of the parameters are as follows: node degrees and community sizes are governed by the power laws, with exponents 2 and 1; the maximum degree is 50; the community size varies between 20 and 100; the mixing parameter $\mu$ varies from 0.1 to 0.3, which is the expected fraction of links of a node connecting it to other communities.

The degree of overlapping is determined by parameters $O_n$ (i.e., the number of overlapping nodes) and $O_m$ (i.e., the number of communities to which each overlapping node belongs).  We fixed the former to be 10\% of the total number of nodes. The latter, the most important parameter for our test, varies from 2 to 8 indicating the diversity of overlapping nodes. By increasing the value of $O_m$, we create harder detection tasks.
 
We compared SLPA with three well-known algorithms, including CFinder (the implementation of clique propagation algorithm \cite{CPM:2005}), Copra \cite{COPRA:2010} (another label propagation algorithm), and LFM \cite{LFM:2009} (an algorithm expanding communities based on a fitness function). Parameters for those algorithms were set as follows. For CFinder, \textit{k} varied from 3 to 10; for Copra, \textit{v} varied from 1 to 10; for LFM $\alpha$ was set to 1.0 which was reported to give good results. For SLPA, the maximum number of iterations \textit{T} was set to 100 and \textit{r} varied from 0.01 to 0.1 to determine its optimal value. The average performances over ten repetitions are reported for SLPA and Copra.
	
\begin{table*}[hbpt]
\centering
\caption{The $Q_{ov}$'s of different algorithms on real-world social networks.}
\label{table:real1}
\begin{tabular}{|c|c|c|c|c|c|c|c|c|c|c|c|} \hline
\textbf{Network} & \textbf{n} & \boldmath{$\overline{k}$} & \textbf{SLPA} & \textbf{std}
& \textbf{r}	& \textbf{Copra}	& \textbf{std}	& \textbf{v}	& \textbf{LFM}	& \textbf{Cfinder}	& \textbf{k}\\ \hline
karate&	34&	4.5&	0.65&	0.21&	0.33&	0.44&	0.18&	3&	0.42&	0.52&	3\\ \hline
dolphins&	62&	5.1&	0.76&	0.03&	0.45&	0.70&	0.04&	4&	0.28&	0.66&	3\\ \hline
lesmis&	77&	6.6&	0.78&	0.03&	0.45&	0.72&	0.05&	2&	0.72&	0.63&	4\\ \hline
polbooks&	105&	8.4&	0.83&	0.01&	0.45&	0.82&	0.05&	2&	0.74&	0.79&	3\\ \hline
football&	115&	10.6&	0.70&	0.01&	0.45&	0.69&	0.03&	2&	 &	0.64&	4\\ \hline
jazz&	198&	27.7&	0.70&	0.09&	0.45&	0.71&	0.05&	1&	 &	0.55&	7\\ \hline
netscience&	379&	4.8&	0.85&	0.01&	0.45&	0.82&	0.02&	6&	0.46&	0.61&	3\\ \hline
celegans&	453&	8.9&	0.31&	0.22&	0.35&	0.21&	0.14&	1&	0.23&	0.26&	4\\ \hline
email&	1133&	9.6&	0.64&	0.03&	0.45&	0.51&	0.22&	2&	0.25&	0.46&	3\\ \hline
CA-GrQc&	4730&	5.6&	0.76&	0.00&	0.45&	0.71&	0.01&	1&	0.45&	0.51&	3\\ \hline
PGP&	10680&	4.5&	0.82&	0.01&	0.45&	0.78&	0.02&	9&	0.44&	0.57&	3\\ \hline
\end{tabular}
\end{table*} 	 
\subsection{Identifying Overlapping Nodes in Synthetic Networks}
Allowing overlapping nodes is the key feature of overlapping communities. For this reason, the ability to identify overlapping \textit{nodes} is an essential component for quantifying the quality of a detection algorithm. However, the node level evaluation is often neglected in previous work. 

Note that the number of overlapping nodes alone is not sufficient to quantify the detection performance. To provide more precise analysis, we define the identification of overlapping nodes as a \textit{binary classification} problem. We use \textit{F-score} as a measure of accuracy, which is the harmonic mean of \textit{precision} (i.e., the number of overlapping nodes detected correctly divided by the total number of detected overlapping node) and  \textit{recall}  (i.e., the number of overlapping nodes discovered correctly divided by the expected value of overlapping nodes, 500 here).

Fig. \ref{fig:FN5000Bmu01} and \ref{fig:FN5000Bmu03} show the F-score as a functions of the number of memberships. SLPA achieves the largest F-score in networks with different levels of mixture, as defined by $\mu$. CFinder and Copra have close performance in the test. Interestingly, SLPA has a positive correlation with $O_m$ while other algorithms typically demonstrate a negative correlation. This is due to the high precision of SLPA when each node may belong to many groups.

\subsection{Identifying Overlapping Communities in Synthetic Networks}
Most measures for quantifying the quality of a partition are not suitable for a \textit{cover} produced by overlapping detection algorithms. We adopted the extended normalized mutual information (NMI) proposed by Lancichinetti \cite{LancichinettiNMI:2009}. NMI yields the values between 0 and 1, with 1 corresponding to a perfect matching. The best performances in terms of NMI are shown in Fig. \ref{fig:NMIN5000Bmu01}
 and Fig. \ref{fig:NMIN5000Bmu03} for all algorithms with optimal parameters.

The higher NMI of SLPA clearly shows that it outperforms other algorithms over different networks structures (i.e., with different $\mu$'s). Comparing the number of detected communities and the average number of detected memberships with the ground truth in the benchmark helps understand the results. As shown in Fig. \ref{fig:framembershipN5000Bmu01}, \ref{fig:rationumcomN5000Bmu01}, \ref{fig:framembershipN5000Bmu03} and \ref{fig:rationumcomN5000Bmu03}, both quantities reported by SLPA are closer to the ground truth than those reported by other algorithms. This is even the case for $\mu$=0.3 and large $O_m$. The decrease in NMI is also relatively slow, indicating that SLPA is less sensitive to diversity of $O_m$. In contrast, Copra drops fastest with the growth of $O_m$, even though it is better than CFinder and LFM on average. 
  
\subsection{Identifying Overlapping Communities in Real-world Social Networks} 
To evaluate the performance of overlapping community detection in real-world networks, we used an overlapping measure, $Q_{ov}$, proposed by Nicosia \cite{Nicosia:2009}. It is an extension of Newman's Modularity \cite{newman-2004-69}. As the $Q_{ov}$ function, we adopted the one used in \cite{COPRA:2010}, $f(x)=60x-30$. $Q_{ov}$ values vary between 0 and 1. The larger the value is, the better the performance is.

In this test, SLPA uses \textit{r} in the range from 0.02 to 0.45. Other algorithms use the same parameters as before. In Table \ref{table:real1}, the \textit{r}, \textit{v} and \textit{k} are parameters of the corresponding algorithms. LFM used $\mu=1.0$. For SLPA and Copra, the algorithms repeated 100 times and recorded the average and standard deviation (std) of $Q_{ov}$. 

As shown in Table \ref{table:real1}, SLPA achieves the highest $Q_{ov}$ in almost all the test networks, except the jazz network for which SLPA's result is marginally smaller (by 0.01) than that of Copra. SLPA outperforms Copra significantly (by $>0.1$) on Karate, celegans and email networks. On average, LFM and CFinder perform worse than either SLPA or Copra. 

To have better understanding of the output from the detection algorithms, we showed (in Table \ref{table:real2})  the statistics, including the number of detected communities (denoted as Com\#) , the number of detected overlapping nodes (i.e., $O_n^d$) and the average number of detected memberships (i.e., $O_m^d$). Due to the space limitation, we present only results from SLPA (Columns 2 to 4) and CFinder (Columns 5 to 7). 

It is interesting that all algorithms confirm that the diversity of overlapping nodes in the tested social networks is small (close to 2), although the number of overlapping nodes differs from algorithm to algorithm. SLPA seems to have a stricter concept of \textit{overlap} and returns smaller number of overlapping nodes than CFinder. We observed that the numbers of communities detected by SLPA are in good agreement with the results from other non-overlapping community detection algorithms. It is consistent with the fact that the overlapping degree is relatively low in these networks. 

\section{Conclusions}
In this paper, we present a dynamic interaction process and one of its implementation, SLPA, to allow efficient and effective overlapping community detection. This process can be easily modified to accommodate different rules (i.e. speaker rule, listening rule, memory update, stop criterion and post-processing) and different types of networks (e.g., k-partite graphs). Interesting future research directions include fuzzy hierarchy detection and temporal community detection.

\begin{table}[hbpt]
\centering
\caption{The statistics of the output from SLPA and CFinder.}
\label{table:real2}
\begin{tabular}{|c|c|c|c|c|c|c|} \hline
   & \multicolumn{3}{c|} {\textbf{SLPA}} & \multicolumn{3}{c|} {\textbf{CFinder}}\\ \hline
\textbf{Network} & \textbf{Com\#}	& \textbf{$O_n^d$}	& \textbf{$O_m^d$}	& \textbf{Com\#}	& \textbf{$O_n^d$}	& \textbf{$O_m^d$}\\ \hline
karate&	2.12&	1.80&	2.00&	3&	2&	2.00\\ \hline
dolphins&	3.44&	1.24&	2.00&	4&	6&	2.00\\ \hline
lesmis&	5.01&	1.13&	2.00&	4&	3&	2.33\\ \hline
polbooks&	3.40&	1.30&	2.00&	4&	9&	2.00\\ \hline
football&	10.30&	1.47&	2.00&	13&	6&	2.00\\ \hline
jazz&	2.71&	2.00&	2.00&	6&	39&	2.05\\ \hline
netscience&	37.84&	2.25&	2.00&	65&	48&	2.33\\ \hline
celegans&	5.68&	14.42&	2.00&	61&	92&	2.70\\ \hline
email&	27.96&	5.57&	2.00&	41&	83&	2.12\\ \hline
CA-GrQc&	499.94&	0.02&	2.00&	605&	548&	2.40\\ \hline
PGP&	1051&	41&	2.00&	734&	422&	2.22\\ \hline
\end{tabular}
\end{table}

\section*{Acknowledgment}

This work was supported in part by the Army Research
Laboratory under Cooperative Agreement Number
W911NF-09-2-0053 and by the Office of Naval Research
Grant No. N00014-09-1-0607. The views and
conclusions contained in this document are those of the
authors and should not be interpreted as representing
the official policies either expressed or implied of the
Army Research Laboratory, the Office of Naval Research, or the U.S. Government.




\bibliographystyle{IEEEtran}  
\bibliography{IEEEabrv,bib/CommunityBIB-Jerry}

\begin{thebibliography}{10}
\providecommand{\url}[1]{#1}
\csname url@samestyle\endcsname
\providecommand{\newblock}{\relax}
\providecommand{\bibinfo}[2]{#2}
\providecommand{\BIBentrySTDinterwordspacing}{\spaceskip=0pt\relax}
\providecommand{\BIBentryALTinterwordstretchfactor}{4}
\providecommand{\BIBentryALTinterwordspacing}{\spaceskip=\fontdimen2\font plus
\BIBentryALTinterwordstretchfactor\fontdimen3\font minus
  \fontdimen4\font\relax}
\providecommand{\BIBforeignlanguage}[2]{{%
\expandafter\ifx\csname l@#1\endcsname\relax
\typeout{** WARNING: IEEEtran.bst: No hyphenation pattern has been}%
\typeout{** loaded for the language `#1'. Using the pattern for}%
\typeout{** the default language instead.}%
\else
\language=\csname l@#1\endcsname
\fi
#2}}
\providecommand{\BIBdecl}{\relax}
\BIBdecl

\bibitem{LancichinettiNMI:2009}
A.~Lancichinetti, S.~Fortunato, and J.~Kert\'esz, ``Detecting the overlapping
  and hierarchical community structure of complex networks,'' \emph{New Journal
  of Physics}, vol.~11, p. 033015, 2009.

\bibitem{CPM:2005}
G.~Palla, I.~Der\'enyi, I.~Farkas, and T.~Vicsek, ``Uncovering the overlapping
  community structure of complex networks in nature and society,''
  \emph{Nature}, vol. 435, pp. 814--818, 2005.

\bibitem{Baumes1:2005}
J.~Baumes, M.~Goldberg, M.~Krishnamoorthy, M.~Magdon-Ismail, and N.~Preston,
  ``Finding communities by clustering a graph into overlapping subgraphs,'' in
  \emph{IADIS}, 2005.

\bibitem{LFM:2009}
A.~Lancichinetti, S.~Fortunato, and J.~Kertesz, ``Detecting the overlapping and
  hierarchical community structure in complex networks,'' \emph{New J. Phys.},
  vol.~11, p. 033015, 2009.

\bibitem{CONGA:2007}
S.~Gregory, ``An algorithm to find overlapping community structure in
  networks,'' \emph{Lect. Notes Comput.Sci.}, 2007.

\bibitem{CONGO:2008}
S.~G., ``A fast algorithm to find overlapping communities in networks,''
  \emph{Lect. Notes Comput. Sci.}, vol. 5211, p. 408, 2008.

\bibitem{COPRA:2010}
{S.. Gregory}, ``Finding overlapping communities in networks by label
  propagation,'' \emph{New J. Phys.}, vol.~12, p. 10301, 2010.

\bibitem{Raghavan:2007}
U.~N. Raghavan, R.~Albert, and S.~Kumara, ``Near linear time algorithm to
  detect community structures in large-scale networks,'' \emph{Phys. Rev. E},
  vol.~76, p. 036106, 2007.

\bibitem{EAGLE:2009}
H.~Shen, X.~Cheng, K.~Cai, and M.-B. Hu, ``Detect overlapping and hierarchical
  community structure,'' \emph{Physica A}, vol. 388, p. 1706, 2009.

\bibitem{ZhangFuzzy:2007}
S.~Zhang, R.-S. Wangb, and X.-S. Zhang, ``Identification of overlapping
  community structure in complex networks using fuzzy c-means clustering,''
  \emph{Physica A}, vol. 374, pp. 483--490, 2007.

\bibitem{Nepusz:2008}
T.~Nepusz, A.~Petr\'oczi, L.~N\'egyessy, and F.~Bazs\'o, ``Fuzzy communities
  and the concept of bridgeness in complex networks,'' \emph{Phys. Rev. E},
  vol.~77, p. 016107, 2008.

\bibitem{NMF:2010}
I.~Psorakis, S.~Roberts, and B.~Sheldon, ``Efficient bayesian community
  detection using non-negative matrix factorisation,''
  \emph{arXiv:1009.2646v5}, 2010.

\bibitem{YYLinkClustering:2010}
Y.-Y. Ahn, J.~P. Bagrow, and S.~Lehmann, ``Link communities reveal multiscale
  complexity in networks,'' \emph{Nature}, vol. 466, pp. 761--764, 2010.

\bibitem{LineGraph1:2010}
T.~Evans and R.~Lambiotte, ``Line graphs of weighted networks for overlapping
  communities,'' \emph{Eur. Phys. J. B}, vol.~77, p. 265, 2010.

\bibitem{Gregory:2010}
S.~Gregory, ``Finding overlapping communities in networks by label
  propagation,'' \emph{New J. Phys.}, vol.~12, p. 103018, 2010.

\bibitem{JieruiXieLPA:2010}
J.~Xie and B.~K. Szymanski, ``Community detection using a neighborhood strength
  driven label propagation algorithm,'' in \emph{IEEE NSW 2011}, 2011, pp.
  188--195.

\bibitem{SteveSurvey:2011}
S.~Gregory, ``Fuzzy overlapping communities in networks,'' \emph{Journal of
  Statistical Mechanics: Theory and Experiment}, vol. 2011, no.~02, p. P02017,
  2011.

\bibitem{LFR:2008}
A.~Lancichinetti, S.~Fortunato, and F.~Radicchi, ``Benchmark graphs for testing
  community detection algorithms,'' \emph{Phys. Rev. E}, vol.~78, p. 046110,
  2008.

\bibitem{Nicosia:2009}
V.~Nicosia, G.~Mangioni, V.~Carchiolo, and M.~Malgeri, ``Extending the
  definition of modularity to directed graphs with overlapping communities,''
  \emph{J. Stat. Mech.}, p. 03024, 2009.

\bibitem{newman-2004-69}
M.~E.~J. Newman, ``Fast algorithm for detecting community structure in
  networks,'' \emph{Phys. Rev. E}, vol.~69, p. 066133, 2004.

\end{thebibliography}

\end{document}